# Validation of Collision Detection and Avoidance Methods for Urban Air Mobility through Simulation

Isha Panchal[1], Sophie F. Armanini[1], and Isabel C. Metz[2]

[1]TUM School of Engineering and Design, Technical University of Munich
[2]Institute of Flight Guidance, German Aerospace Center DLR

**Abstract**

Urban Air Mobility is a new concept of regional aviation that has been growing in popularity as a solution to the issue of ever-increasing ground traffic. Electric vehicles with vertical take-off and landing capabilities are being developed by numerous market companies as a result of the push toward environmentally sustainable aviation. The next stage in the eVTOL development process would be to define the concept of operation of these conceptual aircraft and then to integrate them with the existing airspace once they are airborne. In addition to coordinating with conventional air traffic and other Urban Air Mobility (UAM) vehicles, collision avoidance with uncooperative airspace users has to be addressed. Birds and drones of all sizes could be dangerous for these low-flying aircraft. Innovative collision detection and avoidance techniques need to be employed due to the uncooperative nature of these airspace users and different performance characteristics of urban air mobility vehicles compared to classical fixed-wing aircraft. The aim of this study is to validate one such system by means of fast-time solutions. This system uses a decision tree and safety envelopes to prevent collisions with non-cooperative airspace members. The system is designed to work with different aircraft configurations used for Urban Air Mobility (UAM) operations. Various scenarios are modelled by varying intruder type, location, flight path among others. Changes in flight time and closest point of approach are assessed to evaluate the system with regard to safety and efficiency.

## 1 INTRODUCTION

Thanks to the recent urbanization, most people today live in cities [1]. The sub-urban area has grown dramatically as well. The once nearby villages have now become sub-urban areas as a result of rapid growth of cities. Additionally, there has been a significant increase in travel time [2]. Furthermore, in the future, travel times in ground based transportation systems are expected to grow even longer during peak busy periods, producing unreasonable waiting times [3]. To address this issue, numerous mobility options are being researched. Urban Air Mobility (UAM) is a potential solution that could meet various transportation needs in addition to ground-based transportation [4].

Trying to solve one problem, by adding another segment to the already dense airspace, could lead to more challenges. The UAM aircraft would operate in altitude bands of 2500 ft to 10000 ft with cruise velocities ranging from 110 to 322 km/h [5]. These altitude bands may also contain non-co-operative airspace members (NCOAM) like birds and drone which would not co-ordinate with UAM aircraft and therefore pose a danger to their safe operation [6], [7]. An earlier paper by the authors called for a Collision Avoidance System (CAS) for UAM aircraft considering non-cooperative airspace users [5]. After analysing the regulatory framework and existing CAS, Collision Detect and Resolution (CDR) methods used in conventional aviation, and additionally considering the various configurations of UAM aircraft, a system for typical UAM flight to detect and avoid conflict especially with NCOAMs was proposed [5]. This is termed as Urban Air Mobility Collision Avoidance System (UAM-CAS). The main element is an operational decision tree including measures strategic avoidance prior to flight and measures for tactical/emergency avoidance during flight. The decision tree incorporates different safety envelopes for common UAM configurations as shown in Figure 1.

The decision tree is a collection of procedures to be followed in order to detect and prevent collisions with NCOAM. A flowchart-like tree as seen in Figure 2 is used to represent this set of procedures. This logical structure of the decision tree helps to easily follow the sequence. The fact that the intruder in this study is assumed to be non-co-operative is also taken into account while designing these maneuvers. To analyse and evaluate the efficacy of the UAM-CAS design, a sample flight plan of a shuttle service from an airport to a city centre, representing a typical UAM use-case, was selected [8]. A detailed description can be found in the previous work of the author [5]. At the take-off position, following the decision tree, the aircraft pilot executes the strategic maneuvers and scans the area around the take-off vertiport for suspicious intruder activity before taking off. If there are no intruders, the aircraft can take off. The safety envelopes activate once in the air. The safety envelopes are composed of the caution, warning, and collision envelopes as seen in Figure 1. The encounter with an intruder initiates certain maneuvers from the decision tree when the intruder enters a particular envelope. The tactical maneuver is started when the intruder enters the warning zone. The two phases of the tactical maneuver are DETECT and AVOID. Automated systems carry out collision avoidance during the tactical movement. The human pilot must take control and handle de-escalation if the intruder continues to pose a threat and enters the warning envelope even after completion of the tactical phase. Therefore, with the help of UAM-CAS, conflict with NCOAMs can be prevented and safe operations of the air taxi can be ensured. The envelope's design is based on Aircraft Collision Avoidance System (ACAS) or Traffic Collision Avoidance System (TCAS), a collision avoidance system created for traditional aircraft [9]. Four different configurations of the UAM aircraft are taken into consideration for this study. The UAM-CAS has specially designed envelopes for various configurations due to various flight mechanics resulting from various configurations. The aim of this study is to validate the proposed concept of UAM-CAS by simulations of the maneuvers in the decision tree and the safety envelopes. In order to validate the concept of UAM-CAS and provide tangible results, BlueSky, an open source air traffic simulator, is used [10].

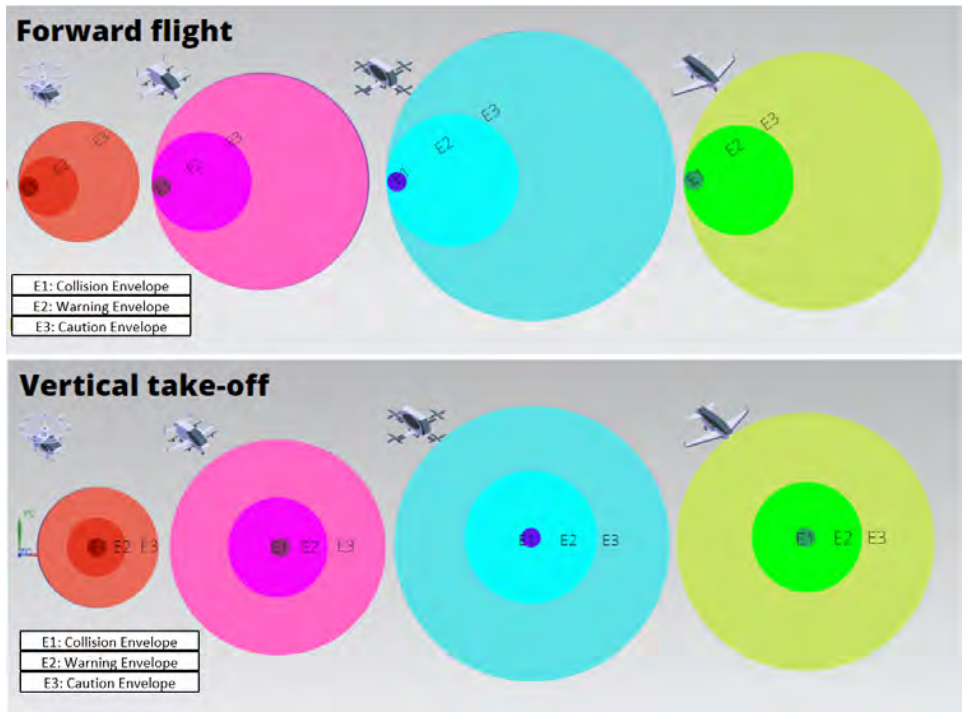

Fig. 1: Collision, warning and caution envelopes for various configurations in forward and vertical flight, source: [5], used with permission

In the upcoming section 2, the methodology to design and build the simulations is explained in detail. The parameters to be evaluated are listed. In the section 3, the obtained results from the simulations for the ground and airborne phase for all four configurations of the ownship are stated and discussed. Finally, in section 4, the results are evaluated and a summary along with the future scope is presented.

## 2    METHODS

In the following sections, details on how the decision tree and safety envelopes are converted into simulation logic are explained. 2.1 provides more details about the simulation environment used. 2.2 shows the breakdown of the various variables and groups it into four different islands. In 2.3, the various phases of the flight and relevant functions which are carried out are mentioned. In the end, section 2.4 lists the parameters chosen to evaluate the results from the simulations.

### 2.1    BlueSky Open Air Traffic Simulator

BlueSky is an open source simulator used for research in domains of air traffic management and to optimise air traffic flow [10].

In order to validate UAM-CAS, plugins containing various sub functions of collision detection and avoidance are developed. For example, functions which facilitate the collision detection and avoidance belong to the CDR plugin. To visualize the aircraft and intruder along with their flight paths, another plugin file executes the logic. The flight path and envelopes are also added as plugins. Similarly, the safety envelopes can be designed and stored in separate plugins. The operator has access to the Anaconda command prompt in addition to the simulation window to start, alter, and finish the simulation. Several statements are printed here during the simulation of the logic . Ultimately, to execute various tests, a collection of commands would be built in the scenario file, and the necessary parameters would be logged.

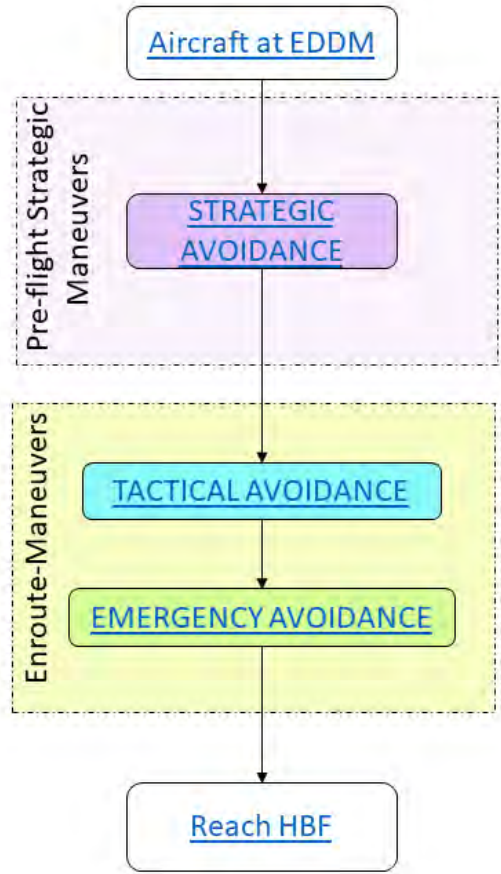

Fig. 2: Decision tree showing the ground and airborne phases of the UAM-CAS, source: [5], used with permission.

For the simulation in BlueSky, the ownship is represented by a yellow arrow with the safety envelopes as concentric circles. The values of the diameter of these circles is assigned in the logic and stated in the plugin file. Once the simulation starts, a pop-up appears on the screen of the operator in the Anaconda command prompt, and they are asked to feed in the type of the ownship. This input is then related to the values of the radii mentioned in the plugin and assigned to the ownship in the current operation. For most of the analysis in this study, configuration 4 - vectored thrust was considered. The caution envelope is yellow in colour, the warning envelope is orange and the collision envelope is red. It is to be noted that, in [5], the author suggested intersecting circles meeting at a common point for forward flight and concentric circles for vertical flight. In contrast to this, in the present work, only concentric circle shaped safety envelopes are considered to ease the simulations. An influence of concentric circle can lead to an increased flight time.

## 2.2 Variables of UAM-CAS : Four Islands

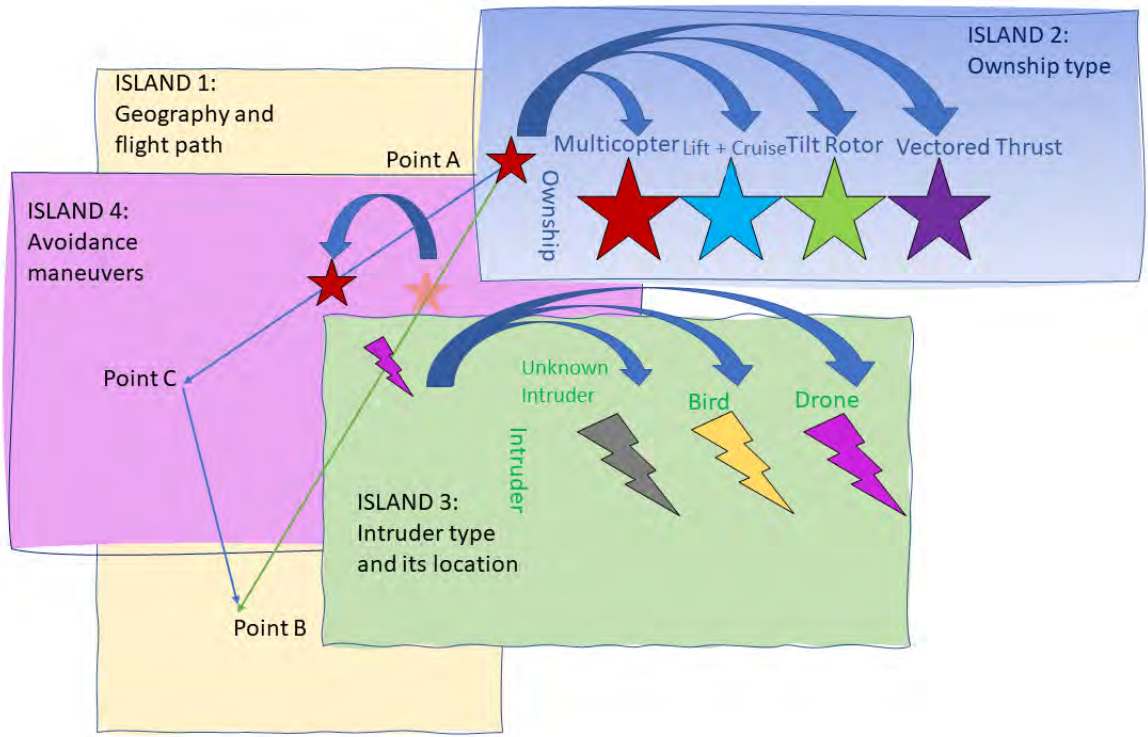

Fig. 3: CAS-UAM subtasks grouped in four islands.

The simulation task in BlueSky can be divided into four topical subtasks called ISLANDS. Island 1 is related to the flight path. Island 2 is related with the ownship. Island 3 deals with intruder and their types. Island 4 contains the collision avoidance maneuvers. Figure 3 shows the islands and their relations [5]. Out of the four islands, three are standalone and can be populated with content without any dependencies on the other islands. Island 4- MANEUVERS, by contrast, overlaps with the other three islands.

1) ISLAND 1- Geography and flight path
   In this island, all parameters related to the geography is taken into consideration. The position of the take-off and the landing vertiports, preferred and alternate routes to be flown, alternate landing vertiport etc., are modelled within this island. This can be done by making a flight related scenario file and then adding stack commands in the code if required.
2) ISLAND 2- Ownship type
   This island deals with the ownship aircraft. All four configurations must be clearly defined. The pre-existing BlueSky aircraft performance model, in this case, is that of an helicopter EC135. The cruise speed, cruise altitude and rate of climb and descent are modified based on the information from the UAM manufacturer [11]–[14]. Based on the configuration, the dimensions of the safety envelope also differ. Moreover, some branches of the tactical collision avoidance are also dependent on the information of the configuration of the ownship. Therefore, a dedicated logic is defined in BlueSky to cater to these requirements.
3) ISLAND 3- Intruder type and its location
   This island deals with the intruder and related factors like its trajectory, location of occurrence and type of intruder . All intruders are NCOAM. The intruder namely, a drone or a bird, is programmed to appear at various locations to initialise and demonstrate the strategic and the tactical maneuver of the mission. The trajectory of the intruder is fed in BlueSky through Comma Separated Values (CSV) files. By doing this, the UAM-CAS can be validated robustly through scenarios.
4) ISLAND 4- Avoidance maneuvers
   This island the collision avoidance maneuvers which are required by the ownship to deviate from the collision course based on logic defined in the decision tree. The incorporation of the required maneuvers within BlueSky is described in section 2.3

Once all these four islands are defined, the individual components can be swapped easily in case the same analysis needs to be performed for any other location or flight characteristics.

### 2.3 UAM-CAS in BlueSky

After implementing all information pertaining to the four islands in BlueSky, the scenarios are defined. Figure 4 shows the detailed flight path, with take-off (V1), landing (V2) and alternate vertiports (V3), along with the ownship, its safety envelopes and the possible location of intruders on route-1 and route-2. The ownship is shown at the vertiport EDDM V1 and its original intention is to fly route 1. NCOAMs are generated in locations 1-4 to test the different branches of the decision tree.

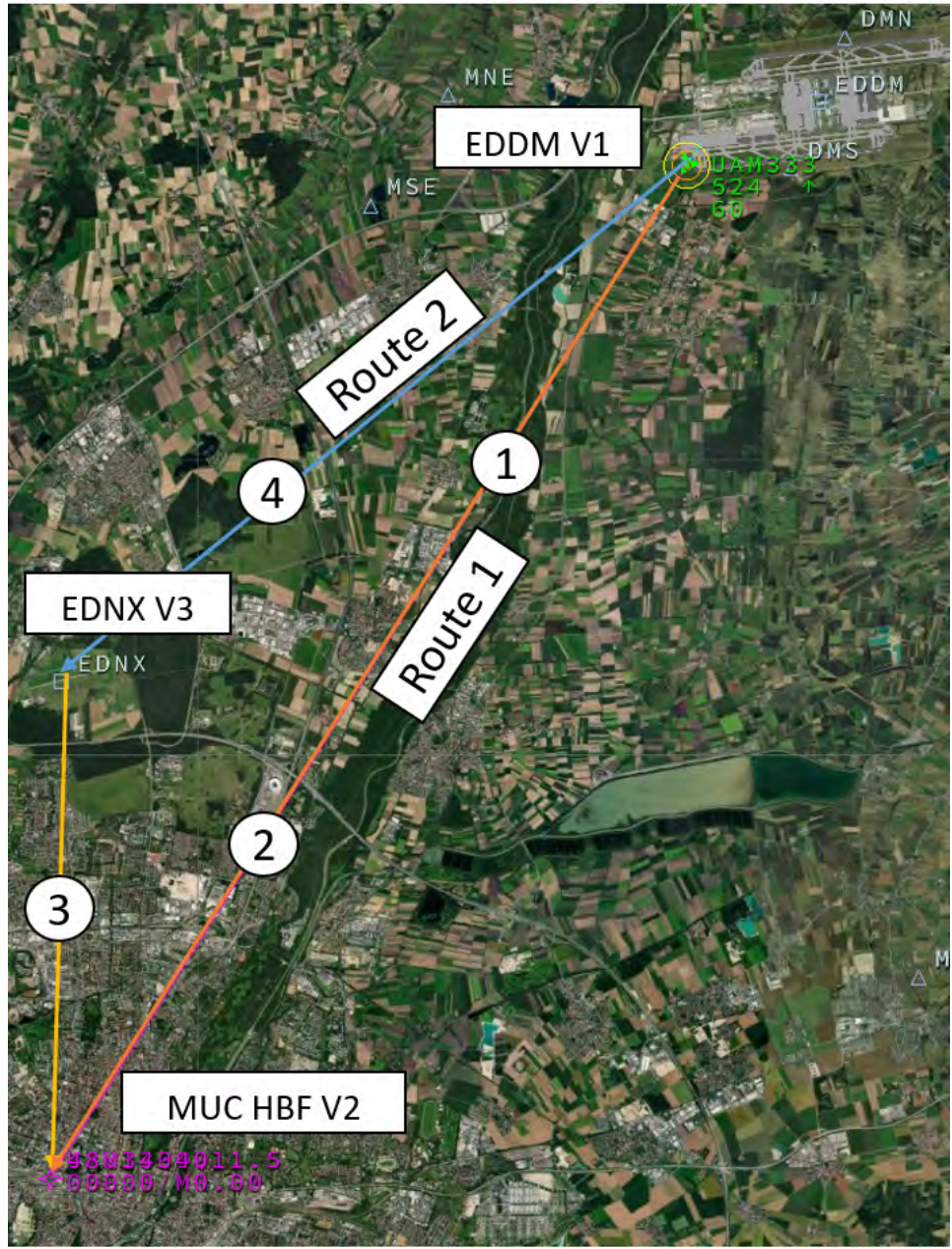

Fig. 4: Vertiports, ownship with safety envelopes, available routes , and location of intruders in scenarios 1-4 on the map view in BlueSky.

Lastly, the maneuvers from the decision trees are implemented. To convert the decision tree into python logic the tree is further broken down into the ground phase, where strategic maneuvers take place, and airborne phase, where tactical and emergency maneuvers are performed, if required.

#### 2.3.1 Ground phase of the decision tree in BlueSky

The ground phase comprises of the strategic avoidance maneuvers, performed by the human pilot.

In BlueSky, when the aircraft is at V1-EDDM, the function related to the strategic maneuver is the Take-off delay check function, which consists of three phases, namely calculation of time of appearance of intruder, separation distance calculation and heading calculation.

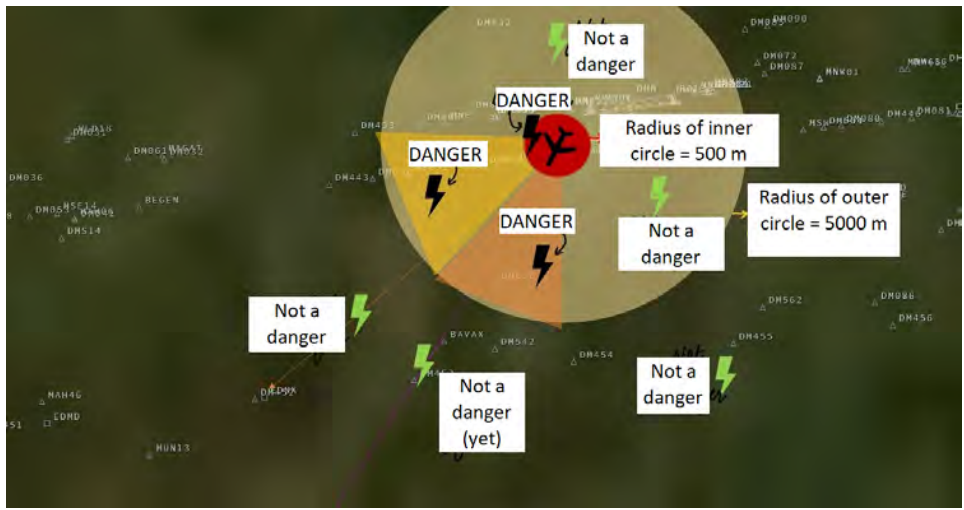

Fig. 5: The various locations of the intruder with respect to the ownship along with indication on whether they pose a danger.

The take-off delay check scans for intruders that might be near the vicinity of take-off vertiport V1 during the time of departure. It filters the intruders on the basis of their heading and separation distance from the ownship. This can be visualised in Figure 5. The green lightning symbol depict intruders which are not headed towards the ownship's flight path whereas the black lightning symbol depict intruders which are headed towards the ownship's flight path in the orange and yellow shaded triangles. An exception to this are those intruders which are extremely near to the aircraft, i.e., within 500 m from V1, in bright-red shaded circular area in Figure 5. These overhead intruders are considered extremely dangerous and are dealt with immediately depending on their heading. Overhead intruders are intruders which might not be heading towards the ownship's flight path but are still present in close vicinity (less than 500 m) of the take-off location V1. One may argue that even intruders which are outside this red shaded area, not headed towards the ownships's planned flight route may suddenly change course and be a danger, but to reduce take-off delays this is omitted in the strategic maneuver and will be dealt with in the tactical maneuver.

### 2.3.2 Airborne phase of the decision tree in BlueSky

The airborne phase consists of the tactical and emergency avoidance maneuvers. The tactical maneuvers are performed by the automated systems on-board the aircraft and if despite them, the conflict is not de-escalated then the human pilot performs the emergency maneuvers. For the scope of this study, the human inputs are also simulated.

- Tactical maneuvers consist of the three elements described below

  1) Ownship configuration check function
     This check determines the type of the ownship from the four possible configurations and then allots the corresponding customised safety envelope to the ownship.

  2) Intruder check function
     With the help of on-board systems and sensors, this function determines the type of the intruder in three seconds of the DETECT phase [5]. This information is very crucial at the later stage for the automated systems to designate tactical avoidance maneuvers in case of impending collision.

  3) Drone direction check function
     In case of an impending collision, the automated systems on-board the aircraft would perform the tactical maneuver to de-escalate the conflict by following the Right of Way (RoW) rules. For this to function as expected, it is very important to know the direction of the drone beforehand. The direction and heading of the drone is detected and displayed on the Anaconda prompt window for situational awareness of the user.

- Emergency Maneuver
  The pilot must conduct an all-systems health check when the conflict has been deemed to be de-escalated and once the ownship is safe. If the ownship has been involved in a collision and sustained physical damage, the pilot must evaluate the situation and take into account all available information on compromised systems. The aircraft batteries may not have enough charge to reach the destination due to the deviation from the intended flight path. Depending on the current position and the aircraft state, the pilot must select whether to return to V1, to continue flying towards the destination vertiport V2 or to divert to the alternate landing location at V3, whichever is nearest, after evaluating all of these data.

|  | Horizontal cruise | Vertical climb | Vertical descent | Theoretical value for total flight |
|---|---|---|---|---|
| Distance on route 1: $d_{R1}(m)$ | 26000 | 304.8 | 304.8 | 26610 |
| Distance on route 2: $d_{R2}(m)$ | 30000 | 304.8 | 304.8 | 30610 |
| Ground speed of vectored thrust ownship : $S_{VT}(m/s)$ | 78 | 1.7 | 1.7 |  |
| Time for route 1: $t_{R1} = d_{R1}/S_{VT}(s)$ | 333.33 | 179.29 | 179.29 | 692 |
| Time for route 2: $t_{R2} = d_{R2}/S_{VT}(s)$ | 384.61 | 179.29 | 179.29 | 744 |

TABLE 1: Theoretical flight values of vectored thrust ownship

This choice will be made after taking into account the distance from the point of conflict to V3. In reality, the pilot may also have to land on a suitable off-airport landing site, but this possibility is not simulated in this study.

### 2.4 Test parameters to be evaluated

To analyse the efficient operation of the UAM-CAS, the following parameters were evaluated in this study.

**Closest point of approach** To evaluate the efficacy of the UAM-CAS, one of the parameters to be analysed from the series of simulations is the Closest Point of Approach (CPA) of the intruder with respect to the ownship. This is the shortest distance between the ownship and the intrusion. In order to be evaluated, this distance is recorded in an output text file.

**Time of simulated flight**

The simulated flight time is a benchmark to evaluate the theoretical time of flight calculated from the data supplied by the UAM manufacturer mentioned in Table 1.

### 2.5 Test scenarios

Various scenarios were tested to validate the different branches of the decision tree. In all scenarios, the ownship is of configuration 4 and the mission is to fly from V1-EDDM to V2-MUC HBF. There are two possible routes: route-1, along the river Isar, and route-2 via V3-EDNX. Intruders like drones and birds can be encountered by the ownship. The theoretical performance data and time of flight for each route is mentioned in Table 1. The vectored thrust ownship is taken as an example. But the same exercise is performed for the other three types of the ownships as well and the results are included in 3. Based on the distance and time values, the theoretical total time of flight is calculated. The theoretical values are taken as the baseline. In the simulation, one reference scenario for each route is performed without any intruders, to calculate the reference total time of simulated flight. Four scenarios for strategic maneuvers are performed to analyse the ground phase of the decision tree.

Fourteen scenarios as mentioned in 2 for tactical and emergency maneuvers with intruders were performed to gather data on simulated flight time and closest point of approach. For example, in Scenario 5, as shown in Figure 6 the route of the ownship, the location of encounter with the intruder, the type of intruder and the behaviour of the intruder are varied. A detailed description of the airborne-phase maneuvers can be seen in Table 2. In the first 13 scenarios, the automated systems along with the human pilot act to resolve the conflict as per the steps mentioned in the UAM-CAS decision tree. In scenario 14, even after the tactical and emergency maneuvers, the conflict is not resolved and the intruder collides with the ownship. This scenario is included to investigate the failure of the UAM-CAS due to unpredictable behaviour of the bird. In the end, the values obtained from all these scenarios for the four different types of ownships are compared in Section 3.

## 3 RESULTS AND DISCUSSION

To evaluate the utility of the proposed UAM-CAS system, two parameters were evaluated namely CPA and total time of simulated flight ($T_S(S)$).

### 3.1 Ground phase

#### 3.1.1 Delay on ground: $D_S(S)$

The ground phase consists of the amount of time the aircraft has spent on the ground at V1, ready for departure. The delay in duration between ready for departure and effective departure is caused by intruder activity in the vicinity of the vertiport. As shown in Table 3, there are five different outcomes possible as discussed in the previous work of the authors [5]. The best case scenario to fly route-1 is at the scheduled departure time. The worst case scenario to fly route-1 is with a delay of five minutes ( or 300 seconds). Similarly for route-2, the best case scenario is to take-off after a six-minute ( or 360 seconds) delay and the worst case is to take-off after a delay of 11 minutes (or 660 seconds). Due to continued presence of intruders, it may also be the case that the departure is called-off. The delay in all the five scenarios can be visualised in Figure 7

As the ownship is not yet moving, the parameter of CPA is not relevant. The delay calculated in Table 3 needs to be added to the delay of the airborne phase to evaluate the final values.

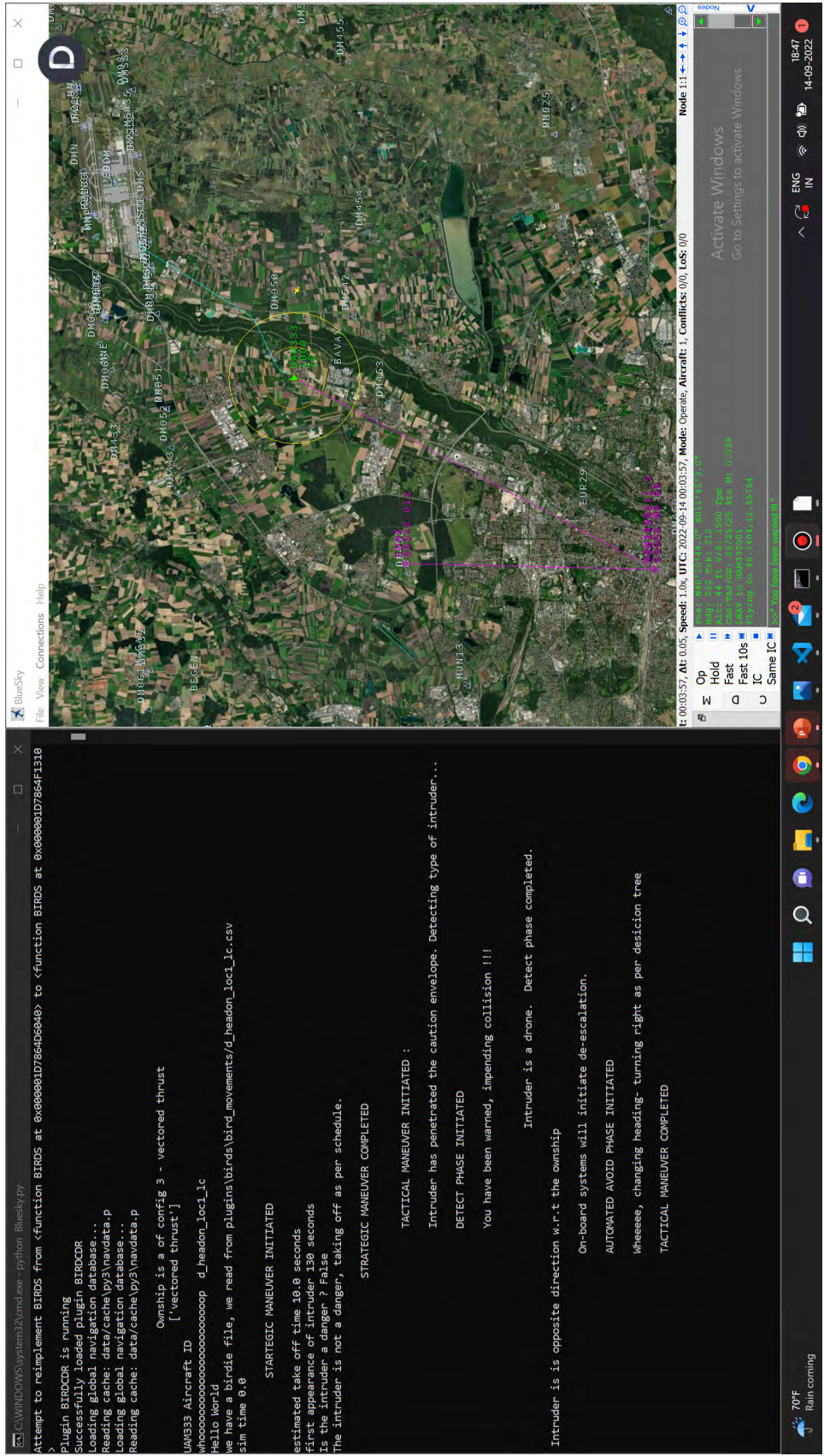

Fig. 6: Simulation screen view with Anaconda command prompt on the left and map view of BlueSky on the right of scenario 5.

| Intruder scenario | | | | | Automated avoid maneuver once intruder enters caution envelope | Emergency maneuver once intruder enters warning envelope |
|---|---|---|---|---|---|---|
| Scenario number | Type | Intent / Behavior | Location of conflict | Direction | | |
| 1 | Drone | predictable | 1 | right | Hover at position, Forward velocity = 0, Drone has RoW | |
| 2 | Drone | unpredictable | 1 | right | Hover at position, Forward velocity = 0, Drone has RoW | Turn left |
| 3 | Drone | predictable | 1 | left | Continue flight, Ownship has RoW | |
| 4 | Drone | unpredictable | 1 | left | Continue flight, Ownship has RoW | Turn right towards V3- EDNX |
| 5 | Drone | predictable | 1 | head-on | Ownship of configuration - 4, turn right | |
| 6 | Drone | unpredictable | 1 | head-on | Ownship of configuration-4, turn right | Continue going right towards V3- EDNX |
| 7 | Drone | predictable | 1 | same direction | If ahead of ownship, drone has RoW, change path | |
| 8 | Drone | unpredictable | 1 | same direction | If ahead of ownship, drone has RoW, change path | Increase separation and continue on parallel path |
| 9 | Bird | predictable | 1 | head-on | Hover and descent to 800 ft, Forward velocity = 0 | |
| 10 | Bird | unpredictable | 1 | head-on | Hover and descent to 800 ft, Forward velocity = 0 | Turn right towards V3-EDNX |
| 11 | Drone | unpredictable | 2 | head-on | Ownship of configuration - 4, turn right | Continue going towards V2- MUC HBF |
| 12 | Drone | unpredictable | 3 | head-on | Ownship of configuration - 4, turn right | Continue going towards V2- MUC HBF |
| 13 | Drone | unpredictable | 4 | head-on | Ownship of configuration - 4, turn right | Continue going towards V3- EDNX |
| 14 | Bird | unpredictable | 1 | head-on | Hover and descent to 800 ft, Forward velocity = 0 | Turn right towards V3-EDNX |

TABLE 2: Description of airborne phase maneuvers for each intruder scenario.

| Possible outcomes | Cause | Delay (sec) |
|---|---|---|
| Scheduled Departure on Route 1 | NCOAM headed to Route 1 or in inner circle | 0 |
| Departure after delay of 5 min on Route 1 | NCOAM stays in vicinity of Route 1 | 300 |
| Departure after delay of 6 min on Route 2 | NCOAM headed to Route 2 | 360 |
| Departure after delay of 11 min on Route 2 | NCAOM stays in the vicinity of Route 2 | 660 |
| Flight is postponed | NCOAM activity has not cleared | infinite |

TABLE 3: Outcome and causes in strategic avoidance in ground phase

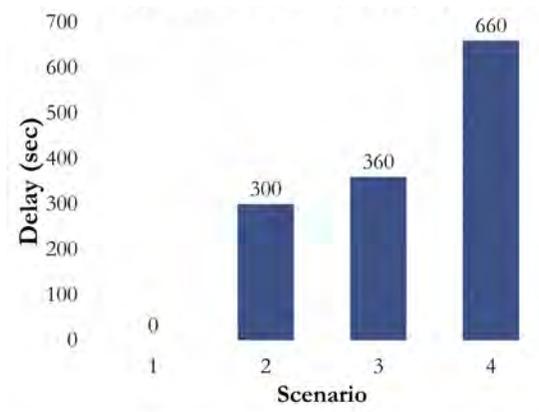

Fig. 7: Delay at the take-off vertiport V1 for each scenario as a part of ground based strategic maneuvers.

### 3.2 Airborne phase
#### 3.2.1 Closest point of approach (CPA)

In Figure 8, the CPA values for scenarios with UAM-CAS and without UAM-CAS are compared for a vectored thrust ownship. It can be observed that except in scenario 10, in all other scenarios, there is a significant drop in CPA. This drop can be as high as 98 %. This reflects that a UAM-CAS is necessary to maintain separation with NCOAMs and without such a system, an intruder could potentially come into dangerous vicinity of the ownship. In Figure 9, it can be observed that in case of unpredictable intruders, the CPA reduces as compared to scenarios with predictable intruders. This is due to the fact that the unpredictable intruders continue getting closer to the ownship and therefore the human pilot has to step in and perform the emergency maneuver once the intruders enter the warning envelope.

| Scenario number | Intruder scenario | | | | Maximum delay in ground phase: $D_G(s)$ | Delay in airborne phase: $D_A(s)$ | Total delay during entire mission: $D_T(s) = D_G + D_A$ |
|---|---|---|---|---|---|---|---|
| | Type | Intent / Behaviour | Location of conflict | Direction | | | |
| 1 | Drone | predicatable | 1 | right | 300 | 531 | 831 |
| 2 | Drone | unpredictable | 1 | right | 300 | 513 | 813 |
| 3 | Drone | predicatable | 1 | left | 300 | 1 | 301 |
| 4 | Drone | unpredictable | 1 | left | 300 | 3 | 303 |
| 5 | Drone | predicatable | 1 | head-on | 300 | 29 | 329 |
| 6 | Drone | unpredictable | 1 | head-on | 300 | 62 | 362 |
| 7 | Drone | predicatable | 1 | same direction | 300 | 64 | 364 |
| 8 | Drone | unpredictable | 1 | same direction | 300 | 60 | 360 |
| 9 | Bird | predicatable | 1 | head-on | 300 | 32 | 332 |
| 10 | Bird | unpredictable | 1 | head-on | 300 | 38 | 338 |
| 11 | Drone | unpredictable | 2 | head-on | 300 | 3 | 303 |
| 12 | Drone | unpredictable | 3 | head-on | 660 | 42 | 702 |
| 13 | Drone | unpredictable | 4 | head-on | 660 | 11 | 671 |
| 14 | Bird | unpredictable | 1 | head-on | 300 | 105 | 405 |

TABLE 4: Values of total delay for entire mission of different scenarios.

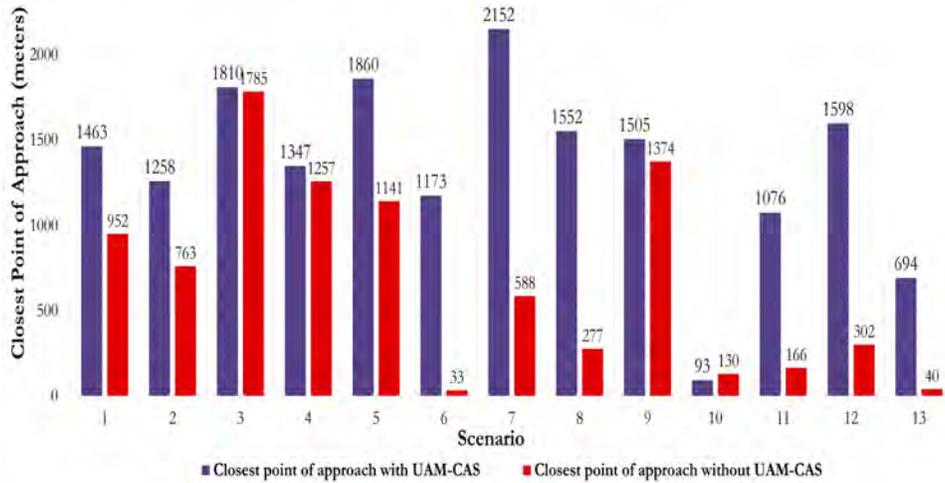

Fig. 8: Closest point of approach with UAM-CAS and without UAM-CAS for each intruder scenario as obtained from the simulations.

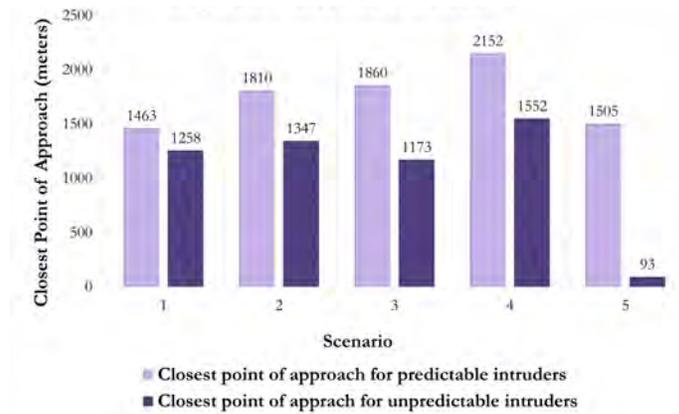

Fig. 9: Comparison of closest point of approach for predictable and unpredictable intruders.

Ownships of type 1 and 2 descended to avoid a head-on collision with drones whereas ownships of type 3 and 4 turned right. Both of these techniques, resulted in successfully evading the intruder. When looking at Figure 10, it can be observed that, CPA has indeed increased for all four types of ownships.

### 3.2.2 Simulated flight time $T_S(S)$ and Delay in Air $D_A(S)$

From Figure 13 for the vectored thrust ownship, it can be observed that the total time of simulated flight for most scenarios (except scenario 1 and 2), is nearly equal to the total time of theoretical flight with no intruders present (maximum difference of 93 seconds). Therefore, by adapting the proposed UAM-CAS as a collision avoidance system, the ownship can maintain safe separation from intruders without prolonging the total flight time substantially. As seen in Figure 12, in scenario 1 and 2, the delay in flight is quite high (maximum difference of 531 seconds). This is due to the fact that the intruder, in this case, a drone entering the safety envelope from the right with respect to the motion of the ownship, lingers in the vicinity of the ownship for a longer time. The strategic avoidance maneuver to avoid collision with this type of an intruder is to hover in place, and therefore, as long as the intruder is in the vicinity of the ownship, the ownship will continue hovering. The time of hover might be a critical factor for Electric Vertical Take Off and Landing (eVTOL) aircraft as the energy required to hover is more than the energy required for forward flight [15], therefore, the avoidance behaviour needs to be optimised. Similarly, as seen in Figure 14, the total time of simulated flight is almost similar or slightly more (maximum difference of 33 seconds) for scenarios with unpredictable intruders as compared to scenarios with predictable intruders.

If the delay is compared for all ownship types, as in Figure 13, can be observed that the delay is the highest for multicopter followed by lift + cruise ones. Ownships of configurations, 1 and 2 transition into hover in case of the threat of head-on collision with a drone. Therefore, the aforementioned increase in the delay could be due to long hover times. The delay is the least for the tilt-rotor configuration of the ownship.

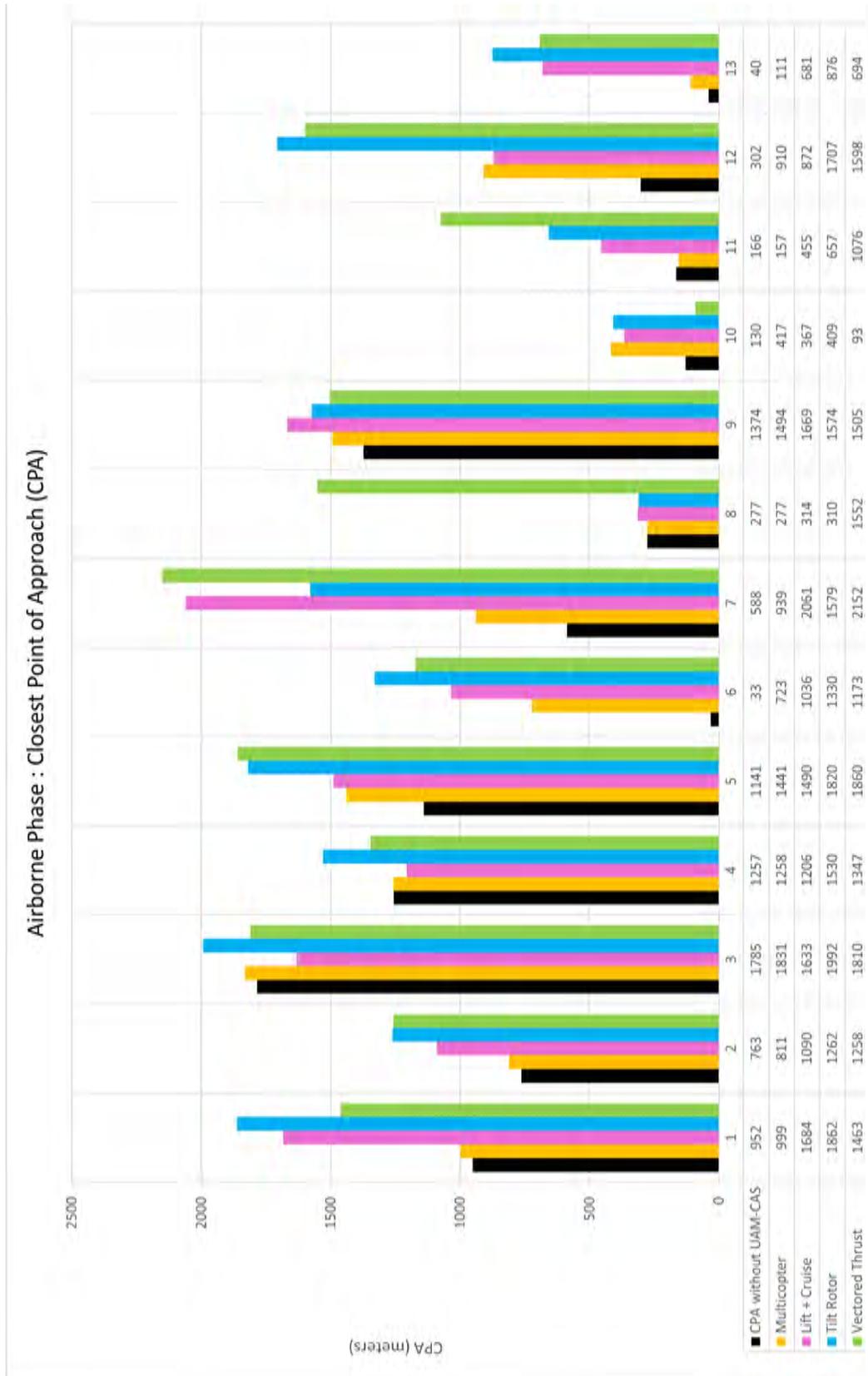

Fig. 10: Comparison of CPA for all four types of ownships.

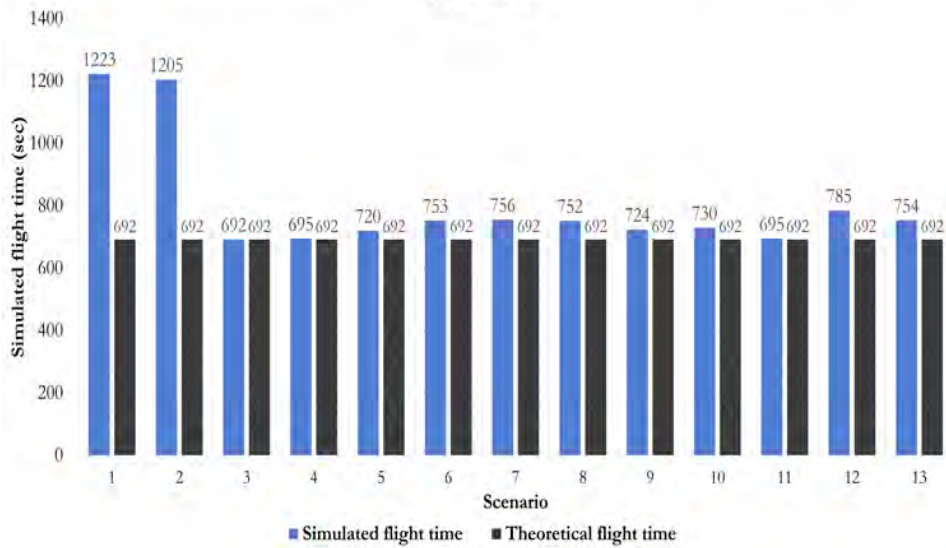

Fig. 11: Total time of simulated flight for each intruder scenario.

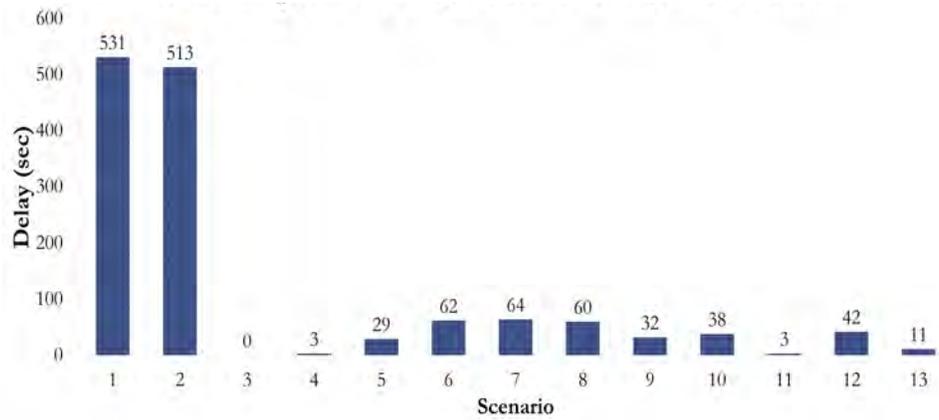

Fig. 12: Delay for each intruder scenario in the airborne phase.

### 3.2.3 Total delay for the entire mission $D_T(S)$

The delay components of the ground phase and the airborne phase were summed up to find the total delay for the entire mission for the vectored thrust ownship. Based on the route, the author decided to calculate the maximum delay for the airport shuttle flight. In scenarios 1-11, the ownship flew on route 1, and as shown in Figure 7, the worst-case delay is five minutes. Therefore $D_G(S)$ is 300 seconds for scenarios 1-11. The ground phase delay is now added to the delay of the airborne phase $D_A(s)$. The maximum total delay $D_T(S)$ is obtained as shown in Table 4. In a similar manner, for route 2, the worst-case ground phase delay $D_G(S)$ is 660 seconds which corresponds to scenario 12 and 13. The values can be visualised in Figure 15. It can be observed that the delay of the ground phase is the larger component and significantly delays the airport shuttle operations. More experimental data would be required to avoid certain time slots to avoid these high delays.

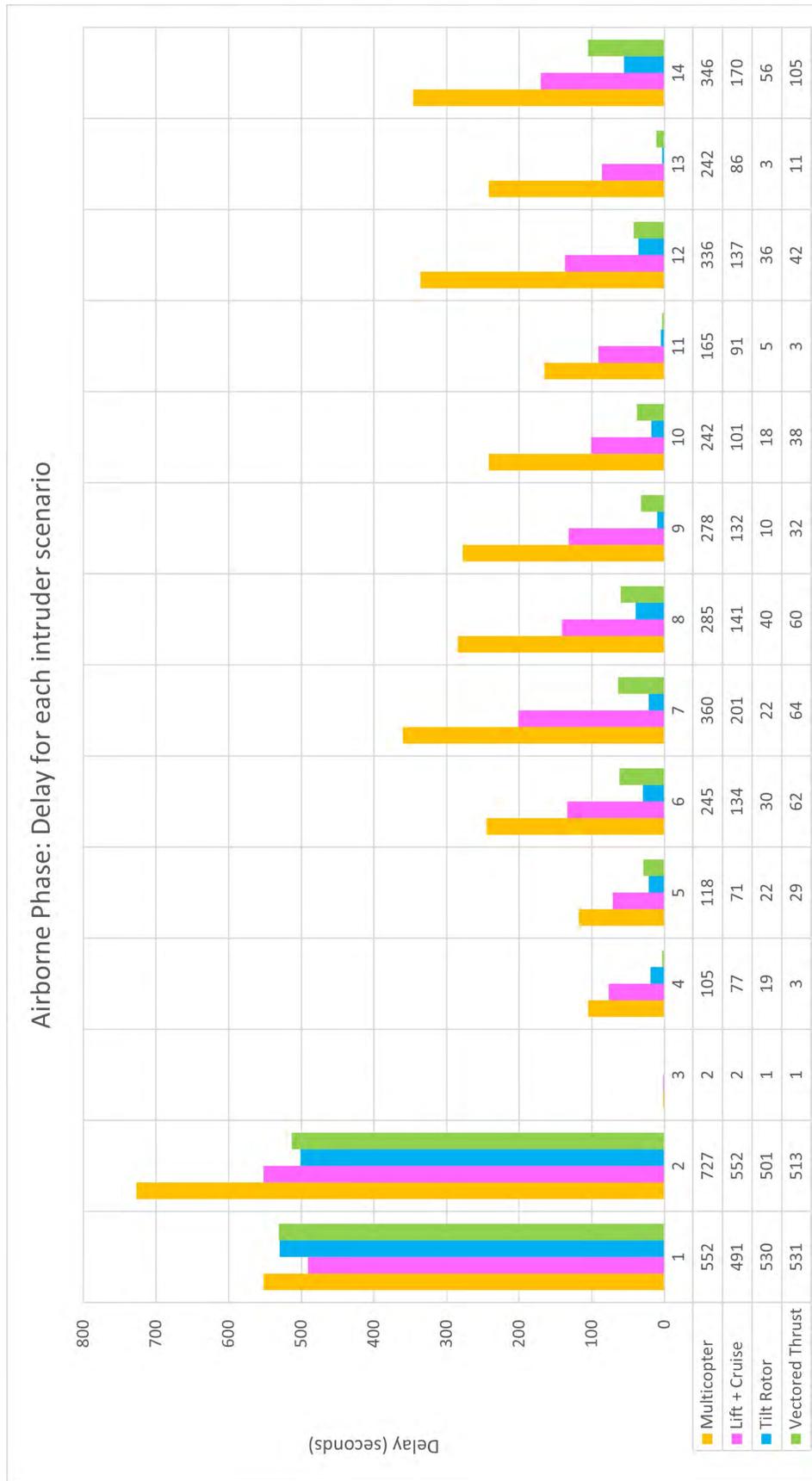

Fig. 13: Comparison of delay in the airborne phase for all four types of ownships.

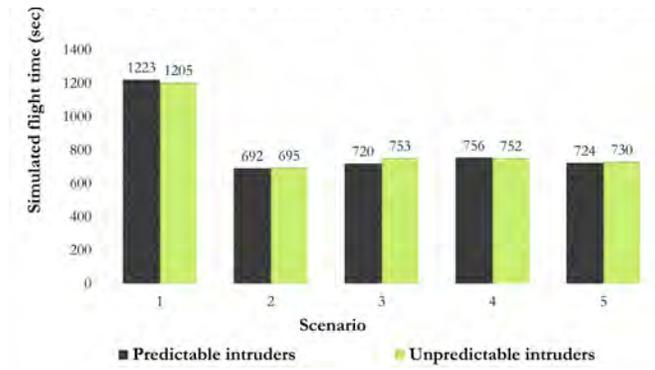

Fig. 14: Comparison of total time of simulated flight for predictable and unpredictable intruders.

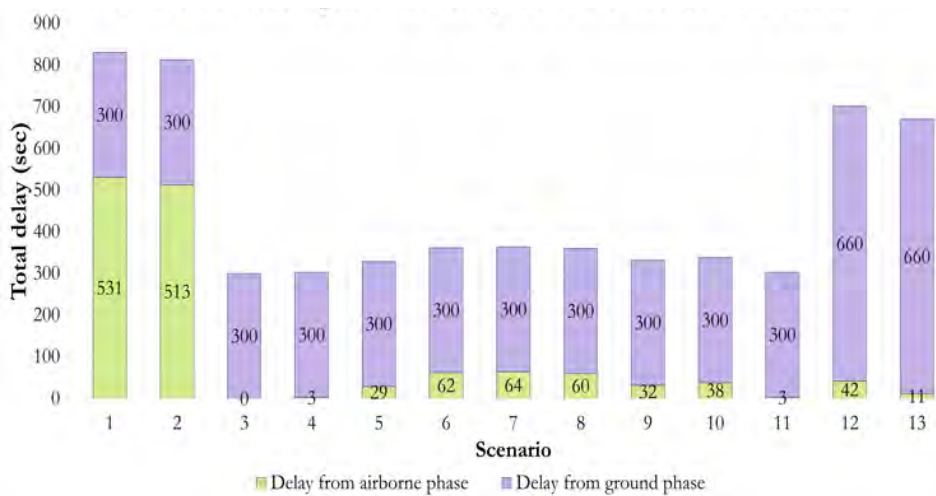

Fig. 15: Maximum total delay comprising of the ground phase component as well as the airborne phase component for each scenario for the entire mission

## 4 CONCLUSIONS

In the United States, the Federal Aviation Administration estimates that flight delays cost airlines 22 billion US Dollars yearly. More than half of the flights in the United States are late (taking account minor delays such as 5 min or 10 min delays) [16], [17]. In Europe, a flight delays cost 100 per minute [18]. This includes everything from extra fuel burn to compensation for passengers.

UAM, which is envisioned as a solution to long queues of on-ground traffic, needs to perform within a tight time schedule to prove its efficiency over conventional solutions. The use-case of an inter-city or intra-city air taxi is largely dependent on the time-saving aspect. Therefore, even delays of a couple of minutes can be a make-or-break scenario. Moreover, with the current generation of electric on-board batteries, the time-of-flight is also limited. These factors make the study of total time of flight and Closest Point of Approach (CPA) worth investigating.

From section 3, it can be concluded that due to assumed delays in the strategic maneuver, the ground phase is responsible for larger delays than the airborne phase. The ground section infrastructure needs to be developed to be at par. Systems like Avian Radio Detection And Ranging (RADAR) could facilitate the detection and monitoring of birds [19], [20]. These systems are not currently widely adopted due to various challenges such as poor detection of birds close to ground, high operating costs, difficulty in integration with the existing Air Traffic Control (ATC) infrastructure, overloading the controllers etc. [21], [22]. This emphasises the need for better surveillance systems and reliable prediction infrastructure. Moreover, to reduce delay involved in the ground phase, shorter waiting times can be adopted after validation through practical experiments. Once the delay resulting from the ground based strategic maneuvers is minimised, the Urban Air mobility Collision Avoidance system (UAM-CAS) can be expected to become an efficient system.

It can also be observed in our results, the decision to hover and wait for a non-cooperative airspace member (NCOAM) to fly-by could leads to longer airborne delay times. With more data, the maneuvers which have high hover-time can be optimised. A workaround in these scenarios could be to re-evaluate the situation after a threshold time and avoid the conflict through change in heading. Although, the manoeuvring capabilities need to be investigated more thoroughly and should be validated again through simulation especially for configuration 1 and 2.

From Figure 10, it can be concluded that by using UAM-CAS, the CPA can be reduced significantly (by upto 72 percentage) below dangerous levels. Moreover as shown in Figure 13, the proposed maneuvers can be successfully carried out without causing excessive delays (average delay of approximately 107 seconds) to aircraft operation. As diversion from planned flight path is not very large, the UAM aircraft can be operated with the current generation of batteries.

The avoidance maneuvers could be further refined based on latest aircraft performance data. By performing further tests with other UAM configurations and with a bigger sample size of scenarios, more reliable results can be achieved. UAM-CAS could act as a basis for a novel Collision Detection and Resolution (CDR) method for electric Vertical Take-off and Landing (eVTOL) flight.

## 5 Funding and/ or Conflicts of interests/ Competing interests

All authors certify that they have no affiliations with or involvement in any organization or entity with any financial interest or non-financial interest in the subject matter or materials discussed in this manuscript.The authors have no competing interests to declare that are relevant to the content of this article.